\documentclass[a4paper,11pt]{article}
\usepackage{soul}
\usepackage[usenames,dvipsnames]{color}
\usepackage{jheppub}
\usepackage{esvect}
\usepackage{amsmath, amssymb, slashed, epsf, color, graphicx, latexsym, tensor, physics, xcolor}
\usepackage{epsfig}
\usepackage{comment}
\usepackage{graphics}
\usepackage{mathtools}
\usepackage{appendix}
\usepackage{inputenc}
\usepackage{indentfirst}

\title{\boldmath Entropy of Flat Space Cosmologies from Celestial dual}

\author[1,2]{Arindam Bhattacharjee,}
\author[3]{Muktajyoti Saha}

\affiliation[1]{Harish-Chandra Research Institute,\\ Chhatnag Road, Jhunsi, Prayagraj—211019, India}
\affiliation[2]{Homi Bhabha National Institute,\\ Training School Complex, Anushaktinagar, Mumbai 400094, India}
\affiliation[3]{Indian Institute of Science Education and Research Bhopal,\\ Bhopal Bypass, Bhauri, Bhopal 462066, India}

\emailAdd{arindamb.hep@gmail.com} 
\emailAdd{muktajyoti17@iiserb.ac.in}

\abstract{We construct a one-dimensional dual theory that effectively describes the sector of the (2+1)D flat gravity phase space near a Flat Space Cosmology (FSC) saddle labeled by definite mass and angular momentum. This Schwarzian type action describes the dynamics of the (Pseudo-) Goldstone Bosons of BMS$_3$ algebra on a circle as the symmetry is spontaneously and anomalously broken. This 1D theory, living on the celestial circle, provides an explicit construction of a celestial dual in (2+1)D. We use it to calculate the semiclassical entropy of Flat Space Cosmologies and find perfect agreement with existing literature.}

\begin{document} 
\maketitle

\section{Introduction}
Understanding the holographic nature of flat spacetime is an intriguing problem. As a generalization of AdS/CFT, it was hinted via holographic reduction in \cite{deBoer:2003vf} that the holographic dual of four-dimensional flat spacetime is a two-dimensional CFT living on the \textit{celestial} sphere. Exploiting the asymptotic symmetries \cite{Strominger:2013jfa, He:2014laa, Kapec:2014opa}, the seminal works \cite{Pasterski:2016qvg, Pasterski:2017kqt} recast the 4D scattering amplitudes written in boost eigenbasis as correlators of a CFT$_2$, dubbed Celestial CFT$_2$. The Ward identities of the currents of the 2D CFT are the soft graviton theorems of the bulk spacetime \cite{Cheung:2016iub, Kapec:2016jld}. The complete description of a celestial dual theory for 4D flat space is however not understood yet. We pose the question for gravity in (2+1)D flat spacetime, which is a much simpler setup due to the absence of local graviton excitations. Here, the nontriviality comes due to the boundary modes associated with the asymptotic symmetries, which in turn dictate the holographic nature. Given that one of the earliest successes of AdS/CFT was to calculate the Bekenstein-Hawking entropy of BTZ black holes in AdS$_3$ using dual CFT description \cite{Strominger:1997eq}, it would be fit to choose 3D flat space as a testing ground for the idea of Celestial Holography.\\

The Poincar\'{e} symmetry of (2+1)D flat spacetime enhances to an infinite dimensional symmetry group near null infinity, called the BMS$_3$ group \cite{Barnich:2010eb}. This group consists of \textit{supertranslations} and \textit{superrotations}, which are infinite dimensional extensions of translations and Lorentz transformations respectively. The corresponding asymptotic symmetry modes span the phase space of (2+1)D gravity with zero cosmological constant. Interestingly, in (2+1) dimensions, there are no asymptotically flat black hole solutions. Nevertheless, there are cosmological solutions called Flat Space Cosmologies (FSC) with non-zero temperature at the horizons \cite{Barnich:2012aw, Barnich:2012xq}.\\

One of the standard approaches towards flat space holography in 3D is to take a large radius limit of AdS$_3$/CFT$_2$. This is dubbed as the BMS$_3$/GCFT$_2$ duality \cite{Bagchi:2009my, Bagchi:2009pe}. GCFTs have a notion of `Cardy regime' and using their own version of modular invariance, the entropy of FSCs were calculated from this dual 2D description \cite{Barnich:2012xq, Bagchi:2012xr}. Another approach is to recast the 3D gravity theory as a 3D Chern-Simons theory, where the dual theory turns out to be a 2D Liouville-like theory \cite{Barnich:2012rz, Barnich:2015sca, Banerjee:2019lrv}. This is based on the duality between Chern-Simons and Wess-Zumino-Witten theories \cite{Elitzur:1989nr, Seiberg:1990eb}. These results suggest that the holographic dual must be a 2D theory lying on the null boundary of 3D flat spacetime. These 2D theories are invariant under the asymptotic symmetry group and are often called BMS field theories. However, this approach seems to be in apparent contradiction with the celestial holography program which in principle should also hold for (2+1)D flat space \footnote{See \cite{Donnay:2022aba,Donnay:2022wvx} for exploration of relations between Carollian and Celestial holographic descriptions for 4D flat spacetimes.}. In fact, it was shown that the BMS$_3$ group has a Virasoro subgroup which coincides with the symmetries of a CFT$_1$ \cite{Barnich:2010eb}. Our major goal would be to address this question and show that indeed (2+1)D asymptotically flat spacetimes can be endowed with a 1D celestial dual structure. This would essentially entail understanding the flat limit of Cardy regime from a 1D perspective and writing down an an effective 1D action whose partition function reproduces the partition function of the bulk gravity theory.\\

In our earlier work \cite{Bhattacharjee:2022pcb}, we have shown the emergence of an effective 1D Schwarzian theory dual to a subspace of (2+1)D Einstein gravity phase space with zero cosmological constant. This subspace is closed under superrotations and its elements are called `Superrotated spacetimes'. The computations rely on foliating these superrotated spacetimes into asymptotically (A)dS$_2$ leaves and use Wedge holography \cite{akal_codimension_2020} to obtain the dual description. The results of \cite{Bhattacharjee:2022pcb} point toward a codimension two holographic nature of gravity in (2+1)D flat space. The superrotation symmetries in 3D form a Virasoro subgroup and clearly the Schwarzian theory breaks the superrotation symmetries. This slight breaking of the full Virasoro group is crucial to get non-trivial dynamics in the boundary Celestial circle. In the works of \cite{Turiaci:2016cvo, Halyo:2019zek}, it has been shown that the (chiral) CFT$_2$ partition function can be understood as a path integral of a Schwarzian theory of Pseudo Goldstone modes corresponding to softly broken Virasoro symmetry due to the presence of anomaly and the finite energy density of the states.\\

In this work, we want to incorporate supertranslations into the picture to describe the full phase space of the theory via a one dimensional dual. Although we could take a geometric approach of foliating these spacetimes with both supertranslations and superrotations along the lines of \cite{Compere:2011ve}, we take a more group theoretic approach \cite{Barnich:2014kra, Barnich:2015uva}. This leads us to identify the (pseudo) Goldstone modes of BMS$_3$ symmetry that non-trivially realises the symmetry on the Celestial circle and write down an effective action for these modes along the lines of \cite{Turiaci:2016cvo, Halyo:2019zek}. 
As we will see, the superrotation part of the dual theory is again a Schwarzian, agreeing with the results of \cite{Bhattacharjee:2022pcb} under an appropriate choice of parameters. \\

The 1D theory is the anticipated celestial dual which effectively describes the dynamics of (2+1)D gravity in a part of the phase space near a flat space cosmological solution. To understand the validity of this proposal, in this work, we compute the semiclassical entropy of 3D Flat Cosmological Solutions from the 1D theory perspective and find agreement with \cite{Barnich:2012xq}. This also implies that our 1D theory is able to capture the Cardy regime of a GCFT$_2$ \cite{Bagchi:2012xr}.\\

The paper is organized as follows: In section \ref{bms3-3d} we discuss the asymptotic symmetries and the phase space of (2+1)D gravity with zero cosmological constant emphasising on cosmological solutions and their thermodynamic properties. We also discuss the lessons learnt from the holographic reduction of superrotated spacetimes. In section \ref{bms3-1d} we discuss the realisation of the BMS$_3$ group on the celestial circle. We write down a celestial dual theory of BMS$_3$ Goldstone modes which effectively describes a sector of the gravity phase space. This is one of the main results of this paper. In section \ref{FSC-thermo} we check the validity of our proposal via the computation of FSC entropy from this dual theory, which agrees with the existing results. Finally, we conclude the paper in section \ref{concl} and discuss possible future directions.

\section{Gravity in (2+1)D and BMS$_3$ symmetry} \label{bms3-3d}

The complete symmetry group of the asymptotically flat 3D spacetimes satisfying flat analogue of Brown-Henneaux type boundary conditions is called BMS$_3$ group. For completeness, we present here a brief introduction to the phase space of (2+1)D gravity and how the algebra arises there as the asymptotic symmetry algebra.

\subsection{Phase space of flat 3D gravity}
We begin by writing the 3D metric in Bondi gauge and then demanding that the metric satisfy 3D Einstein equations with zero cosmological constant. This gives the general form of the metric to be \cite{Barnich:2010eb},
\begin{align}
    ds^2 = \mathcal{M}(\theta)du^2 - 2dudr + (u\mathcal{M}'(\theta) + 2\mathcal{J}(\theta))dud\theta + r^2 d\theta^2.
\end{align}
where $\mathcal{M}(\theta)$ and $\mathcal{J}(\theta)$ are arbitrary functions. Then we identify the vector fields that leave the form of the above metric invariant upto leading order. These are called asymptotic Killing vectors and for the above class of metric they are given by \cite{barnich_dual_2013},
\begin{align}\label{killing-bms3}
\chi^u &= T(\theta) + u Y'(\theta), \nonumber\\
\chi^{\theta} &= Y(\theta) - \frac{1}{r} \partial_\theta\chi^u, \nonumber\\
\chi^r &= -r \partial_{\theta}\, \chi^{\theta} + \frac{1}{2r}(uM'(\theta)+J(\theta)) \partial_{\theta}\chi^u , 
\end{align}
where again we have $T(\theta)$ and $Y(\theta)$ as arbitrary functions that parameterise the gauge transformations. The function $T(\theta)$ parameterises an abelian vector field and it is called supertranslation as it generalises spacetime translation to angle dependent transformations. On the other hand, the vector fields parameterised by $Y(\theta)$ generalises global Lorentz transformations and are called superrotation vectors.\\

Since these are Killing vectors, we can calculate the charges associated with these vectors. For a given spacetime labelled by ($\mathcal{M}(\theta),\mathcal{J}(\theta)$) they are given by \cite{barnich_dual_2013}, 
\begin{align}\label{bms-charges}
Q_{\chi}[\mathcal{M},\mathcal{J}] = \frac{1}{16\pi G} \int d\theta (\mathcal{M} T + \mathcal{J} Y).
\end{align}
It is interesting to observe that although $\mathcal{M}$ appears in the supertranslation part of the conserved charge above, starting from the flat space solution $\mathcal{M} = -1/8G$ and $\mathcal{J} = 0$, a generic superrotation $g \xrightarrow{} g + \mathcal{L}_{Y(\theta)} g$ would generate non-trivial angle dependent $\mathcal{M}(\theta)$\footnote{However, not all of these solutions are valid spacetimes in the gravity phase space as they may contain conical deficits \cite{Barnich:2012aw}.}. Thus we will call these spacetimes `superrotated spacetimes' even though they are characterised by supertranslation charges.\\

These charges, when endowed with a proper Dirac Bracket, satisfy the 3D BMS algebra or $\mathfrak{bms}_3$. Written in terms of the modes of $\mathcal{M}$ and $\mathcal{J}$, the algebra take the following form,
\begin{align}
    &[L_m,L_n] = (m-n)L_{m+n} + \frac{c_1}{12}(m^3-m)\delta_{m+n,0} 
 \nonumber\\
    & [L_m,M_n] = (m-n)M_{m+n} + \frac{c_2}{12}(m^3-m)\delta_{m+n,0} \nonumber \\
    & [M_m,M_n] = 0. \label{bms3-algebra} 
\end{align}
For pure gravity, $c_1 = 0$ and $c_2 = 3/G$. Nonzero $c_1$ arises if we add a Lorentz Chern-Simons term along with pure gravity. The global part of this algebra spanned by the $m = 0,\pm 1$ modes coincide with the 3D Poincar\'{e} algebra.\\

Since pure gravity in (2+1) dimensions does not contain local degrees of freedom, i.e. gravitational waves, the only non triviality comes from the boundary. Hence the functions $\mathcal{M}(\theta)$ and $\mathcal{J}(\theta)$ span the phase space of the theory. On the other hand, the vector field labelled by $T(\theta)$and $Y(\theta)$ transforms the elements of this phase space to each other. These two sets of functions can then be thought of as elements of dual spaces where the inner product is given by the surface charge (\ref{bms-charges}).

\subsection{Superrotated spacetimes and their dual theory} \label{superrotated-review}

In an earlier work \cite{Bhattacharjee:2022pcb}, we have considered the part of the phase space consisting of superrotated spacetimes. The general form of these spacetimes is given by,
\begin{align}\label{superrotated-metric}
    ds^2 = \mathcal{M}(\theta)du^2 - 2dudr + 2u\mathcal{M}'(\theta) dud\theta + r^2 d\theta^2.
\end{align}
Choosing an origin in the bulk, we may slice up these spacetimes into foliations of fixed proper time (distance). The region inside the past and future light cone of the origin in this co-ordinate looked like warped product of (A)AdS$_2 \ltimes \mathbb{R}$, 
\begin{align}
    ds^2 = -d\tau^2 + \tau^2\left[d\tilde{\rho}^2 + \left(\frac{1}{4}\text{e}^{2\tilde{\rho}} -\frac{\mathcal{M}}{2} + \frac{\mathcal{M}^2}{4}\text{e}^{-2\tilde{\rho}} \right) d\theta^2\right]. \label{Bondi_in_hyper}
\end{align}
where (A)AdS$_2$ stands for asymptotically anti de Sitter spacetimes in 2D in the sense of \cite{grumiller_menagerie_2017}. Similarly, the spacetime outside of the lightcone has an asymptotically dS$_2 \ltimes \mathbb{R}$ geometry. It is straightforward to see that the introduction of supertranslation breaks this warped product structure, at least in the leading order. Although if we allow for terms upto a certain order in the mixed components, a similar structure may emerge as in \cite{Compere:2011ve}.\\

Next, we consider a wedge region in the future lightcone bounded by two (A)AdS$_2$ surfaces at $\tau_1$ and $\tau_2$ and perform a dimensional reduction of the pure gravity theory inside the wedge onto these boundaries via wedge holography \cite{akal_codimension_2020, ogawa_wedge_2022, Bhattacharjee:2022pcb}. The idea is to eventually cover the full spacetime by taking $\tau_1 \xrightarrow{} 0$, $\tau_2 \equiv \tau_{\infty} \xrightarrow{} \infty$ and also by doing the same procedure for dS slices. We find that the massless modes of the dimensionally reduced theory give us pure gravity on the boundary (A)AdS$_2$ slices. This action is non-dynamical and proportional to the Euler character of the manifold. This led us to the insight that if superrotation symmetry is unbroken, then there is no dynamics in this smaller subspace of phase space.\\

To get a non-trvial dynamics, we break the superrotation symmetry by fluctuating the (A)AdS$_2$ branes along the transverse direction $\tau_i$. This fluctuation in the position of the brane is sourced by a scalar mode $\phi$ as $\tau_i \xrightarrow{} \tau_i (1+\phi_i)$. Now performing the dimensional reduction and some change of variables, the effective theory on (A)AdS$_2$ slices turns out to be the JT-gravity where the $\phi$ mode plays the role of the dilaton
\begin{align}
    S_{\text{eff}} =& \frac{\tau_\infty}{16\pi G_N} \left(\int d^2x \sqrt{g}[ R + \phi(R+2)] + \int dx\sqrt{\gamma}\phi(K-1) \right). \label{Seff_weyl}
\end{align}
Crucially the boundary term of the 3D gravity action consistent with our phase space also matches with the boundary term of the JT gravity action on AdS slices.\\

Further, we can write down a 1D dual action by integrating out the dilaton field in JT action which acts as a Lagrange multiplier. The resulting action (adding both AdS and dS slice contributions) is a Schwarzian theory \cite{Bhattacharjee:2022pcb} where the bulk field $\mathcal{M}(\theta)$ plays the role of the Schwarzian derivative of the time reparameterisation mode(as we can show $\mathcal{M}(\theta) = \{Y(\theta),\theta\}$), 
\begin{align}
    S_{1D} = \frac{\phi_r}{8\pi G_2} \int d\theta \mathcal{M}(\theta). \qquad G_2 \equiv \frac{G_N}{\tau_\infty}, \quad \phi_r \xrightarrow{} \text{boundary value of dilaton}  \label{1D_eff}
\end{align}

The insight we get from this is that the bulk superrotation symmetry has a non-linear realisation in terms of Goldstone modes on the Celestial circle. Thus to generalise this result for the full phase space of the 3D  gravity theory, we need to understand the effective field theory of Goldstone modes corresponding to the asymptotic symmetries. Around a classical solution of Einstein's equations, this effective theory would govern the semiclassical dynamics and would describe an explicit Celestial holographic dual description. Below we describe the bulk solutions in brief detail along with their thermodynamic properties. We will see these solutions and their properties emerge from the explicit 1D dual that we would construct later.

\subsection{Flat space cosmologies}
Among these asymptotically flat metrics, we will be particularly interested in Flat Space Cosmologies (FSC). (2+1)D gravity in flat space does not have black hole solutions but these are the closest analogues of black holes in 3D flat space. Flat Space Cosmologies are parameterized by $\mathcal{M}(\theta)= 8GM$ and $\mathcal{J}(\theta) = 8GJ$, where $M > 0$ is the mass and $J$ is the angular momentum. Thus in BMS gauge, these solutions take the following form \cite{Barnich:2012aw, Barnich:2012xq},
\begin{align}
    ds^2 = 8GMdu^2 - 2dudr + 8GJdud\theta + r^2 d\theta^2.
\end{align}
Observe that the usual flat space can be obtained by putting $\mathcal{M}(\theta) = -1/8G$ and $\mathcal{J}(\theta) = 0$. Also, for the BTZ black holes, the angular momentum $J$ is restricted for given $M$ which is not the case for these cosmological solutions. In the phase space of 3D gravity, the flat space is disconnected from the FSC solutions \cite{Barnich:2012aw}, much like the spectrum of BTZ black hole from global AdS$_3$. The FSC solution has a cosmological horizon, located at
\begin{align}
    r = r_C = \sqrt{\frac{2GJ^2}{M}}.
\end{align}
This horizon is also a Killing horizon, generated by the vector $K = \partial_u + \Omega\partial_\theta$. The FSC can be associated with angular velocity and inverse temperature given by,
\begin{align}
    \Omega = -\frac{2M}{J}, \quad T = \beta^{-1} = \frac{2}{\pi}\sqrt{\frac{2GM^3}{J^2}}.
\end{align}
The thermodynamics\footnote{Although there are subtleties with some signs in the first law, as compared to standard black hole thermodynamics.} of these solutions are well-studied in the literature \cite{Barnich:2012xq}. The Bekenstein-Hawking entropy is given by,
\begin{align}
    S_{\text{FSC}} = \frac{2\pi r_C}{4G} = \frac{\pi\abs{J}}{\sqrt{2GM}}. \label{FSC-ent}
\end{align}
In \cite{Barnich:2012xq}, it was shown that the thermodynamics of these solutions can have a holographic interpretation similar to BTZ black holes in AdS$_3$. Here, the entropy can be derived from a suitable limit of the Cardy formula in CFT. Such computations rely on an appropriate modular invariance of a CFT$_2$ partition function on a torus. This was also computed from a 2D perspective in \cite{Bagchi:2012xr} along the line of BMS$_3$/GCFT$_2$ correspondence. Logarithmic correction to the semiclassical entropy has also been calculated in \cite{Bagchi:2013qva}. To make this connection, the FSC solution is analytically continued to the Euclidean signature. The demand of a smooth geometry fixes the periodicity of the Euclidean time direction to $\beta$. Whereas, the periodicity of the angular direction gets fixed to $i\beta\Omega = i\Phi$, typical for rotating black holes \cite{Barnich:2012xq}. \\

The main aim of this paper is to compute the semiclassical entropy from an independent 1D celestial dual theory.

\section{BMS$_3$ from 1D perspective}\label{bms3-1d}
To understand and construct the 1D dual theory we would extensively use the alternate representation of $\mathfrak{bms}_3$ algebra in terms of vector fields on a circle. In this section, we will briefly describe that structure, which is extensively studied in \cite{Barnich:2014kra, Barnich:2015uva, Oblak:2016eij}. It is evident looking at \eqref{bms3-algebra} that the superrotation subalgebra of $\mathfrak{bms}_3$ is just Virasoro algebra. So we begin by understanding how Virasoro algebra arises on a circle and how we can recreate the Schwarzian action discussed in \cite{Bhattacharjee:2022pcb} purely from a 1D perspective.

\subsection{Virasoro group on a circle}
From the bulk perspective it is evident that the effect of superrotation on a circle at constant $u$ at null infinity is to induce a reparameterisation of $\theta$, the co-ordinate that parameterises the circle. Hence we consider the group of orientation preserving diffeomorphisms (or conformal transformations) on S$^1$. The transformations are specified by the group of smooth functions $f$ such that,
\begin{align}
    \text{Diff}(S^1) = \{f:S^1\rightarrow S^1 \,|\, f(\theta + 2\pi) = f(\theta) + 2\pi, \, f'(\theta)> 0\}. 
\end{align}
The corresponding Lie algebra is spanned by infinitesimal diffeomorphisms $f(\theta) = \theta + Y(\theta)$, which are generated by vector fields. It is given by,
\begin{align}\label{virasoro-vectors}
    \text{Vect}(S^1) = \{Y(\theta)\partial_{\theta} \,|\, Y(\theta + 2\pi) = Y(\theta)\}.
\end{align}
It can be seen that the basis elements $l_m = \text{e}^{im\theta}\partial_{\theta}$ satisfy the Witt algebra with respect to the standard Lie bracket
\begin{align}
[l_m, l_n] = (m-n)l_{m+n}.
\end{align}
The corresponding Lie group is Diff(S$^1$). Now since we are interested in the quantum Hilbert space and the action of the symmetry group on it, we would be mostly interested in the projective representations of the symmetry group instead of the exact one. Thus, we would look at the possible central extensions of the Diff(S$^1$) group. Of course, this has a non-trivial central extension, where it coincides with the Virasoro group, given by:
\begin{align}
    [l_m,l_n] = (m-n)l_{m+n} + \frac{c}{12} m(m^2-1)\delta_{m+n,0}. \label{virasoro}
\end{align}
Since we have already seen the circle equivalent of superrotations as diffeomorphism generating vector fields, let us now try to understand how the charges of the form (\ref{bms-charges}) appear from this perspective. For this, we would need to understand what would be the dual space of (\ref{virasoro-vectors}). We would follow the notations and conventions of \cite{Oblak:2016eij} for most of this discussion.\\

We begin by considering the generalisation of vectors, which are \textit{densities} on a circle and defined by their transformation properties under diffeomorphism. A density $\alpha$ of weight $h$ transforms\footnote{In presence of a central term, the transformation will be modified nontrivially but we will skip the details here. We will write the final result for BMS$_3$ as that will be relevant for our computations. For details, see \cite{Oblak:2016eij}.} in the following way under a diffeomorphism:
\begin{align}
    (f\cdot\alpha)(f(\theta)) = \frac{\alpha(\theta)}{(f'(\theta))^h}. \label{diff-on-density}
\end{align}
which in analogy with CFT literature transforms like a conformal primary of weight $h$. The vector space of $h$-densities is denoted as $\mathcal{F}_{h}(S^1)$ and the transformation law (\ref{diff-on-density}) gives an infinite dimensional representation of Diff(S$^1$) on this vector space. The vector fields of (\ref{virasoro-vectors}) are densities of weight $-1$. Since this space of vector fields is also the Lie algebra of the Virasoro group, the action (\ref{diff-on-density}) for $h=-1$ is the adjoint action of the group.\\

Now the charges would be the diffeomorphism invariant quantities constructed out of these densities. For this we would need the dual of the vector space $\mathcal{F}_{h}(S^1)$ whose elements would map $\alpha \in \mathcal{F}_{h}(S^1)$ to $\mathbb{R}$,
\begin{align}
    p: \alpha \longmapsto \langle p,\alpha\rangle \equiv \frac{1}{2\pi} \int_0^{2\pi} d\theta p(\theta)\alpha(\theta). \label{Noether}
\end{align}
By demanding invariance under diffeomorphism, i.e $\langle f\cdot p,\alpha\rangle = \langle p,f^{-1}\alpha\rangle$ we can show that $p \in \mathcal{F}_{1-h}(S^1)$. Thus we conclude that the dual space of densities of weight $h$ are densities of weight $1-h$. Indeed for vectors ($h = -1$), the duals are quadratic densities with $h =2$. Keeping up with the nomenclature, the action of the group on this dual vector space is called the coadjoint action. From (\ref{diff-on-density}) we see that their infinitesimal transformation law is, 
\begin{align}
    Y\cdot p = Yp' + 2p Y'. \label{alg-on-density} 
\end{align}
This is of course recognisable as the infinitesimal transformation law of stress tensors of a usual 2D CFT. Thus we see that the charges are bilinears constructed out of the vectors and their duals.\\

The above discussion was restricted to Diff(S$^1$) and not its centrally extended counterpart, the Virasoro group. To get the adjoint and coadjoint action of the Virasoro group from that of Diff(S$^1$) we would also need the knowledge of the non-trivial cocycles of the group. Instead of going into the details, let us just state the results here.\\
Every element of the Virasoro group is a pair $(f,a)$ where $f$ is an element of Diff(S$^1$) and $a \in \mathbb{R}$. The corresponding Lie algebra also gets a nontrivial central extension. The adjoint action of this group on its Lie algebra elements (given by $(Y,\lambda)$) is,
\begin{align}
    Ad_{f} (Y, \lambda) = (f\cdot Y, \lambda - \frac{1}{12} \langle S[f],Y \rangle).
\end{align}
Note that, the central term $a$ of the group element labeling the adjoint transformation does not appear in the transformation law. We see that the first term transforms exactly as a vector under Diff(S$^1$) while the second term gets a shift proportional to the quadratic density $S[f]$, which is the Schwarzian derivative of $f$ with respect to $\theta$, defined as,
\begin{align}
    S[f](\theta) = -\frac{1}{2}\left(\frac{f''}{f'}\right)^2 + \left(\frac{f''}{f'}\right)'.
\end{align}
The coadjoint action also changes similarly. The elements of the coadjoint space are now also paired $(p,c)$ where $p$ is a quadratic density and $c \in \mathbb{R}$. The coadjoint action of the group is then given as
\begin{align}
    Ad^*_{f} (p,c) = (f\cdot p -\frac{c}{12} S[f^{-1}],c ).
\end{align}
Here we see the effect of central term on the transformation properties of our `stress tensor'. Written in a more familiar form, for a conformal transformation $\theta \xrightarrow{} \xi(\theta)$, the transformation law above takes the well known expression,
\begin{align}
\tilde{T}(\xi(\theta)) = (\xi'(\theta))^{-2} T(\theta) - \frac{c}{12} S[\xi^{-1}](\theta).
\end{align}
Thus we see that the familiar CFT stress tensor has a 1D analogue in terms of an element of the coadjoint space. This observation further extends below where we understand the dynamics of a chiral CFT via Goldstone modes on a 1D circle.

\subsection{BMS$_3$ on the circle}
Now that we understand the realisation of superrotation (Virasoro) part of the asymptotic symmetry algebra on a circle, we want to also realise the supertranslation part. From the bulk perspective, supertranslation induces an angle dependent translation along $u$ direction, $u \xrightarrow{} u + \alpha (\theta)$. If we think of the celestial circle as a circle at constant $u$, then clearly this transformation induces a field in the circle. The induced field must be abelian as the effect of two supertranslations (in absence of any superrotations) is just additive $u \xrightarrow{} u + \alpha (\theta) + \beta(\theta)$.\\

A much more non-trivial transformation occurs when both supertranslation and superrotations are present. Under a superrotation (or diffeomorphism) a supertranslation changes in the following way,
\begin{align*}
    \alpha (\theta) \xrightarrow{f(\theta)}  f'(\theta) \alpha (\theta),
\end{align*}
which from (\ref{diff-on-density}) we know is the transformation property of vector fields ($h=-1$). Hence, we know that supertranslations induce vector fields of the form $\alpha(\theta) \partial_{\theta}$ on the celestial circle. \\

The BMS$_3$ group is a semidirect product of the group of superrotations (which are the diffeomorphisms on the circle) and the abelian group of vector fields corresponding to supertranslations. This structure is that of an \textit{exceptional} semidirect product since the second component of the product is the Lie algebra of the first component. The vector fields are acted on by the elements of the diffeomorphism group according to the adjoint representation. Of course, now we can think of the centrally extended BMS$_3$ group whose elements are going to be labelled as $(f,\lambda; \alpha,\mu)$ where $(f,\lambda)$ constitutes the superrotation subgroup and $(\alpha,\mu)$ constitutes the supertranslation subgroup. $\lambda$ and $\mu \in \mathbb{R}$ are the central extensions. The infinitesimal transformations are generated by the elements of the corresponding Lie algebra $\mathfrak{bms}_3$. In this case, the generators of both supertranslation and superrotation are vector fields. Thus the corresponding currents are quadratic densities, which are elements of the dual space of the Lie algebra. \\

Just like Virasoro, here also we would look at the space of quadratic densities for the coadjoint action of the BMS$_3$ group. This space is spanned by two quadratic densities $(j,p)$ and two central elements $(c_1,c_2)$. Under finite BMS$_3$ transformation labelled by\footnote{Note that we have suppressed the central terms $(\lambda, \mu)$ since they do not appear in the transformations.} $(f,\alpha)$, the coadjoint element $(j,c_1;p,c_2)$ transforms to $(\tilde{j},c_1;\tilde{p},c_2)$ such that \cite{Oblak:2016eij},
\begin{align}
    p(\theta) &= [f'(\theta)]^2\Tilde{p}(f(\theta)) - \frac{c_2}{12}S[f](\theta), \label{p-trans}\\
    j(\theta) &= [f'(\theta)]^2\Tilde{j}(f(\theta)) - \frac{c_1}{12}S[f](\theta) \nonumber \\
    & + f'(\theta) \left[\alpha\circ f (\tilde{p}\circ f)' + 2(\alpha\circ f)' \tilde{p}\circ f \right](\theta)  -\frac{c_2}{12}[f'(\theta)]^2 \alpha'''\circ f(\theta). \label{j-trans}
\end{align} 
Let us look at the transformation properties closely. The supertranslation parameter $\alpha$ does not appear at all in the transformation \eqref{p-trans} of the supertranslation charge (or \textit{supermomentum}) $p(\theta)$. Under pure superrotation i.e. for $\alpha = 0$, both the charges transform like the holomorphic component of 2D stress tensor. Note that, if we restrict ourselves to the superrotation subsector of the asymptotic symmetries, we only get nontrivial transformation for the supertranslation charge in pure Einstein gravity for which $c_1 = 0$. This feature will be extremely important when we will point out the connection between the 1D dual theory obtained here and the results of \cite{Bhattacharjee:2022pcb}.

\subsection{(Pseudo) Goldstone bosons of BMS$_3$}
Our discussion about the coadjoint representations of BMS$_3$ group is motivated by the fact that we want to write an action on the celestial circle that mimics the dynamics of Einstein Gravity on the bulk. Let us begin by identifying the states in the Hilbert space of this dual theory.\\

Since $(p(\theta), j(\theta))$ are the stress tensor equivalents for BMS$_3$, we should label the states on our Hilbert space via the mutually commuting zero modes of these two fields. This precisely follows \cite{Bagchi:2009pe} where a similar description comes from considerations of Galilean conformal algebra, $\mathfrak{gca}_2$. Thus the states are labelled as $|h_j,h_p \rangle$ such that,
\begin{align}\label{state-labels}
    \langle j(\theta) \rangle = h_j, \qquad \langle p(\theta) \rangle = h_p.
\end{align}
These eigenvalues are closely related to the mass and angular momentum of the bulk flat space cosmology that these states correspond to. To see this, observe that from (\ref{j-trans}) the modes of the fields $(p(\theta), j(\theta))$ satisfy $\mathfrak{bms}_3$ algebra almost identical to (\ref{bms3-algebra}) apart from the central term,
\begin{align}
       & [j_m,j_n] = (m-n)j_{m+n} + \frac{c_1}{12}m^3 \delta_{m+n,0} \nonumber \\
       & [j_m,p_n] = (m-n)p_{m+n} + \frac{c_2}{12}m^3 \delta_{m+n,0} \nonumber \\
       & [p_m,p_n] = 0.
   \end{align}
By shifting the zero modes i.e. $(j_0, p_0)$ appropriately $j_0' = j_0 + \frac{c_1}{24}$ and $p_0' = p_0 + \frac{c_2}{24}$, we can bring this algebra into the form of (\ref{bms3-algebra}). Since mass $M$ and angular momentum $J$ are defined via the zero modes of (\ref{bms3-algebra}), they are also similarly related.
\begin{align}\label{relating-mj-h}
    J = h_j + \frac{c_1}{24}, \qquad M = h_p + \frac{c_1}{24}.
\end{align}

Now suppose we have a state labelled by $(h_j,h_p)$ on the dual theory. For large charges, this can actually be thought of as an ensemble labelled by a finite temperature and a chemical potential. The finite temperature part is easy to understand if we think of the superrotation charge $j(\theta)$ as the stress tensor of the theory whereas we think of $p(\theta)$ as an abelian spin 2 current giving rise to the chemical potential. This finite energy density of these states breaks the BMS$_3$ symmetry spontaneously into the global symmetry group. BMS symmetry is also anomalous due to the presence of the central charge. Thus we expect that the symmetry is non-linearly described in this part of the Hilbert space via the pseudo Goldstone Bosons. In other words, we expect a dynamical theory of these modes, where the BMS$_3$ symmetry is softly broken in the action and the breaking is governed by the central charges. This idea imitates \cite{Turiaci:2016cvo} where the authors describe the holomorphic sector of a CFT near a finite energy state in terms of pseudo Goldstone modes. Below we extend that idea from that of the Virasoro algebra to $\mathfrak{bms}_3$ algebra.\\

We assume that the eigenvalues (\ref{state-labels}) are large enough compared to the central charges. With this assumption let us start from our chosen state and do a small superrotation and superstanslation labelled by $f(\theta)$ and $\alpha(\theta)$ respectively. Then according to the expressions (\ref{p-trans}) and (\ref{j-trans}), the expectation values would change accordingly.
\begin{align}
    \langle p(\theta) \rangle =& h_p\, [f'(\theta)]^2 - \frac{c_2}{12} S[f](\theta),\label{exp-p-trans}\\
    \langle j(\theta) \rangle =& h_j\, [f'(\theta)]^2 - \frac{c_1}{12} S[f](\theta) - \alpha (f(\theta))\left( 2f''(\theta) h_p - \frac{c_2}{12}\frac{S[f]'(\theta)}{f'(\theta)}\right).\label{expj--trans}
\end{align}
The homogeneous terms in these formulae shows exactly how spin 2 fields transform under a diffeomorphism. The central terms are present due to conformal anomaly and oblivious to the supertranslation. The final term in (\ref{expj--trans}) is proportional to the supertranslation and is obtained from (\ref{j-trans}) upto some total derivative factors. Since we are in a compact manifold (circle), we may omit those total derivative terms which will be inconsequential in our later considerations.\\

The main idea here adopted from \cite{Turiaci:2016cvo} is to interpret this $f(\theta)$ and $\alpha(\theta)$ not as co-ordinate transformation and gauge parameters\footnote{We will see that a particular combination of these BMS$_3$ transformation parameters can be thought of as inducing gauge transformations.} but rather as local quantum fields. In fact, these are the Goldstone bosons of BMS$_3$ that non-linearly realises the symmetry. Since we have assumed $h_p, h_j \gg c_1, c_2$, we can fairly assume that the state $\ket{h_p,h_j}$ has a large degeneracy. This is essentially analogous to the ``Cardy regime'' of CFTs. Then if we zoom into the part of the Hilbert space near the state $\ket{h_p,h_j}$ (denoted by $\mathcal{H}_{j,p} \subset \mathcal{H}$), we can promote (\ref{exp-p-trans}, \ref{expj--trans}) to operator identities.
\begin{align}
     p(\theta)|_{\mathcal{H}_{j,p}}  =& h_p\, [f'(\theta)]^2 - \frac{c_2}{12} S[f](\theta),\label{p-operator}\\
     j(\theta)|_{\mathcal{H}_{j,p}} =& h_j\, [f'(\theta)]^2 - \frac{c_1}{12} S[f](\theta) - \alpha (f(\theta))\left( 2f''(\theta) h_p - \frac{c_2}{12}\frac{S[f]'(\theta)}{f'(\theta)}\right).\label{j-operator}
\end{align}
Since these are now operator statements and we have identified $p(\theta)$ and $j(\theta)$ as the spin 2 current and the stress tensor of our theory respectively, we can write down an effective action on the celestial circle,
\begin{align}
       I = &-\frac{\beta}{2\pi}\int_0^{2\pi} d\theta\left([f'(\theta)]^2 h_p - \frac{c_2}{12}S[f](\theta) \right) \nonumber \\
       &-\frac{\Phi}{2\pi}\int_0^{2\pi} d\theta \left[(f'(\theta))^2 h_j - \frac{c_1}{12}S[f](\theta) - \gamma(\theta)\left( 2f''(\theta) h_p - \frac{c_2}{12}\frac{S[f]'(\theta)}{f'(\theta)}\right)\right] \label{bms-PGB}
   \end{align}
where $\gamma \equiv \alpha\circ f$ acts as a Lagrange multiplier. The coupling constants $\beta$ and $\Phi$ are still arbitrary but they will soon be related to bulk quantities as suggested by the notations.\\

We propose that the above action captures the gravitational dynamics near a flat space cosmological solution. As a consistency check, let us return to the case of purely superrotated spacetimes in pure Einstein gravity. For these spacetimes $h_j = 0 = c_1$ and also $\gamma$ is absent. Thus the above action reduces to,
\begin{align}
       I_{s} = -\frac{\beta}{2\pi}\int_0^{2\pi} d\theta\left([f'(\theta)]^2 h_p - \frac{c_2}{12}S[f](\theta) \right). \label{superrotation-PGB}
   \end{align}
By rescaling the field $f(\theta) \xrightarrow{} \sqrt{\frac{24h_p}{c_2}} f(\theta)$ and then by defining $Y(\theta) = \tanh(f(\theta)/2)$ we get,
\begin{align}
       I_{s} = \frac{\beta}{8\pi G}\int_0^{2\pi} d\theta \{Y(\theta), \theta\}.
   \end{align}
we exactly get the form of (\ref{1D_eff}) back given that the parameter $\beta$ and the dilaton boundary value $\phi_r$ are related appropriately\footnote{To be precise, we need to identify the combination $\tau_{\infty} \phi_r$ with $\beta$, where $\tau_{\infty}$ is a cutoff as discussed in section \ref{superrotated-review}.}.\\

In the next section we use the action (\ref{bms-PGB}) to calculate the bulk thermodynamic properties providing further evidence of its validity.

\section{Properties of the effective action}\label{FSC-thermo}
In the last section, we have written a 1D theory on the celestial circle. Our claim is that the corresponding path integral is equivalent to the 3D gravity path integral around an FSC saddle. To check the validity of this statement, our preliminary goal would be to understand whether we can map 3D gravity solutions in this 1D theory language under suitable choice of parameters. In particular, we would like to obtain the entropy of flat space cosmologies (FSC), which are labeled by mass $M$ and angular momentum $J$.\\

In 3D gravity picture, the gravity path integral is performed on a grand canonical ensemble. The ensemble is labelled by the inverse temperature $\beta$ and the angular velocity $\Phi$ which are the conjugate variables to energy and angular momentum respectively. In the thermodynamic limit, they are related to the FSC parameters $M$ and $J$. To understand these relations from the perspective of our 1D theory, we need to understand the vacuum orbit representative of the theory for a given $(\beta, \Phi)$.\\

We know that BMS$_3$ is the group of diffeomorphisms and vector fields on the circle on which the group of diffeomorphisms acts with the adjoint action. Now the quadratic densities $(p(\theta), j(\theta))$ form an element of the coadjoint vector space of BMS$_3$ group. The coadjoint vector space can be foliated into subspaces that are completely disjoint from each other and that are closed under the operation of the group. These are called co-adjoint orbits. Our state (\ref{state-labels}) is a representative of one such orbit.\\

Let us start with the assumption that as $\Phi = 0, \beta \xrightarrow{} \infty$ the orbit representative is the state $\ket{h_p = 0, h_j = 0}$ and then we would try to evaluate the representative state at a finite value of the parameters. In order to do so, locally we will think of BMS$_3$ as the following set of infinitesimal transformations
\begin{align}
    \theta \to& \theta + \epsilon(\theta),\nonumber \\
    x \to& x + g(\theta); \qquad g(\theta) \equiv \frac{\alpha(\theta)}{\epsilon(\theta)} \label{gauge-trans}
\end{align}
where $\theta$ is the co-ordinate along the circle on which diffeomorphism acts. The interesting point to note is that the transformation of the parameter $x$ depends only on the coordinate of the circle i.e. we have an angle-dependent translation along the $x$ direction. To understand the implications of such an auxiliary direction, we can start from the usual 2D notion of BMS$_3$ group where it transforms the coordinate along null infinity $u \to u + u \epsilon'(\theta) + \alpha(\theta)$ and then define $x$ via  $u \equiv \epsilon(\theta) x$. Note that we have already discussed that both $\alpha (\theta)$ and $\epsilon(\theta)$ are generated by vector fields on S$^1$. Hence, the transformation properties of $g(\theta)$ dictate that it is a function or a density of weight $h = 0$. Admittedly, $g(\theta)$ is not well defined when $\epsilon(\theta) \to 0$ but since we will be concerned about finite transformations here, we may overlook it for this context. The effective 1D description comes because the transformation parameters depend only on the coordinate on the circle and the $u$-dependence is trivial. Hence, all the data available at a particular constant $u$-slice can be trivially evolved to other constant $u$-slices. Thus from a Kaluza-Klein perspective, we can 
\textit{exactly} interpret the 2D coordinate transformations as coordinate and gauge transformations in 1D. The transformations \eqref{gauge-trans} exactly do this job. Now we can think of the direction $x$ to be trivially parameterising the fiber on the base manifold S$^1$ and null infinity can essentially be thought of as the principal bundle.  \\

Then we can consider a finite version of the above transformations by exponentiating the infinitesimal ones,
   \begin{align}
       & x \to x + G(\theta), \qquad G(\theta) = \frac{\alpha}{\epsilon}\circ f (\theta), \nonumber \\
       & \theta \to f(\theta).
   \end{align}
To go to the finite parameter saddle from the 
$h_j=h_p=0$ vacuum, we would like to put certain periodicity conditions on the circle and the fiber \cite{Detournay:2012pc, Afshar:2019tvp}. Thus we have the following BMS$_3$ transformations,
\begin{align}
       f(\theta) = \text{e}^{\frac{2\pi\theta}{\Phi}}, \quad G(\theta) = -\frac{\beta}{\Phi}\theta, \implies \alpha\circ f(\theta) = -\frac{2\pi\beta}{\Phi^2}\theta f(\theta),
\end{align}
where we get the identification ${(\theta,x)\sim (\theta + i\Phi, x - i\beta)}$ and the conditions $f(\theta + i\Phi) = f(\theta) + i\Phi$ and $G(\theta + i\Phi) = G(\theta)$. We can put these forms back in (\ref{j-trans}) and (\ref{p-trans}) to get the following relations,
\begin{align}
         h_p = \frac{c_2\pi^2}{6\Phi^2}, \quad h_j = \frac{c_1\pi^2}{6\Phi^2} - \frac{c_2\pi^2}{3}\frac{\beta}{\Phi^3}.  \label{parameters-relation}
\end{align}
For a cosmological solution of $M, J \gg c_2$, and for pure gravity $c_1 = 0$. So, from above equation and (\ref{relating-mj-h}) we get,
\begin{align}
         M = \frac{c_2\pi^2}{6\Phi^2}, \quad J = - \frac{c_2\pi^2}{3}\frac{\beta}{\Phi^3}.
\end{align}
These match with \cite{Barnich:2012xq} and also justify our notation of the parameters $\beta$ and $\Phi$ as inverse temperature and angular velocity. Having correctly identified the 3D saddles from a 1D perspective, we will now show the computation of the thermodynamic variables using the 1D theory \eqref{bms-PGB}, when we identify the couplings with these $\beta$ and $\Phi$.

\subsection{Entropy of flat space cosmologies}
With the information of the coadjoint orbits at hand, we may now write down the path integral of the 1D theory formally as,
   \begin{align}
       Z = \int \mathcal{D}f \mathcal{D}\alpha\ \text{e}^{-I}.
\end{align}
with $I$ given by (\ref{bms-PGB}) with the couplings identified with the `inverse temperature' and `angular velocity' parameters obtained in the earlier part of this section. To evaluate this partition function completely would require us to understand the measure in the phase space. We leave this complete analysis for a later work and instead focus here on a semiclassical computation.\\

In the large charge limit, the path integral can be approximated by saddle point contributions only. For this, we would consider the solutions to the equations of motion of the action \eqref{bms-PGB}. The Lagrange multiplier field $\gamma = \alpha\circ f$ puts a constraint on $f$ given as,
\begin{align}
    2f''(\theta) h_p - \frac{c_2}{12}\frac{S[f]'(\theta)}{f'(\theta)} = 0.
\end{align}
The solutions to this equation are of the form $f = A\theta$. The solution for $\gamma$ comes from the equation of motion of the field $f$. For the linear solutions of $f$, the corresponding $\gamma$ must satisfy the following equation,
\begin{align}
    2h_p\gamma''(\theta) - \frac{c_2}{12A^2}\partial^{4}_{\theta}\gamma(\theta) = 0.
\end{align}
However, the explicit form of $\gamma$ does not appear in the construction. Under saddle point approximation, the partition function is given by the onshell action for $f = A\theta$ such that,
   \begin{align}
       \log{Z} = - I_{\text{onshell} } = - A^2 \frac{c_2\pi^2}{6}\frac{\beta}{\Phi^2} + A^2\frac{c_1\pi^2}{6\Phi}.
   \end{align}
From the definition of charges i.e. $-\partial_{\beta}\log{Z} = h_p $ and $-\partial_{\Phi}\log{Z} = h_j$, we fix the constant $A = 1$. Therefore, we have identified a solution in the 1D theory that is the representative of the FSC solution in 3D gravity.\\

Now we consider the entropy that is given by the Laplace transform of partition function\footnote{The unusual sign of the entropy is typical for FSC solutions \cite{Barnich:2012xq}.},
   \begin{align}
       \text{e}^{-S(h_j,h_p)} = \int d\beta d\Phi \text{e}^{\beta h_p + \Phi h_j} Z(\beta,\Phi).
   \end{align}
Under saddle point approximation, we have $-S(h_j,h_p) = \beta h_p + \Phi h_j + \log{Z}$, where $\beta, \Phi$ take the thermodynamic values. Using \eqref{parameters-relation} and $h_p\approx M, h_j\approx J$, we get the semiclassical entropy,
   \begin{align}
       S = \frac{\pi}{\sqrt{6}}\left(\sqrt{\frac{c_2}{M}}J + c_1\sqrt{\frac{M}{c_2}}\right).
   \end{align}
   Our result is in perfect agreement with the GCFT computation in \cite{Bagchi:2012xr, Bagchi:2013qva}. For pure gravity, we have $c_1 = 0$ and $c_2 = 3/G$, such that
   \begin{align}
       S = \pi \frac{\abs{J}}{\sqrt{2GM}}.
   \end{align}
   This agrees with the semiclassical entropy of FSC \cite{Barnich:2012xq}.

\section{Discussions}\label{concl}

In this work, we have found a 1D theory \eqref{bms-PGB} of pseudo Goldstone modes corresponding to the spontaneously broken BMS$_3$ symmetry. This theory lies on the celestial circle of flat space. To arrive at this theory, we have considered the transformation properties of supertranslation and superrotation charges under the action of finite BMS$_3$ transformations. We have identified a solution in this 1D theory that represents the 3D FSC solution with definite mass and angular momentum. Using saddle point approximation to compute the path integral corresponding to the 1D theory, we have obtained the semiclassical entropy for this 1D solution. This perfectly agrees with the Bekenstein-Hawking entropy of the 3D FSC \cite{Barnich:2012xq} in Einstein gravity. The 1D theory also captures the Cardy formula of a GCFT$_2$ \cite{Bagchi:2012xr}.

\subsection*{Comparison with holographic reduction}

In an earlier work \cite{Bhattacharjee:2022pcb}, we obtained a Schwarzian theory on the celestial circle describing the dynamics of superrotation modes. We arrived at this theory via holographic reduction of 3D Einstein theory to Jackiw-Teitelboim (JT) gravity on a cutoff (A)dS surface and eventually using JT-Schwarzian duality. We observed that the boundary theory is an integration of the supermomentum parameter, which is a Schwarzian derivative of the reparametrization mode of the boundary. This theory also depends on two arbitrary parameters $\tau_{\infty}$ and $\phi_r$. Here $\tau_{\infty}$ is the choice of cutoff surface and $\phi_r$ governs the boundary behavior of the fluctuation in the location of this cutoff surface. As pointed out earlier, from \eqref{p-trans} we understand how supermomentum is blind to the supertranslation transformations. To compare with pure Einstein theory we consider $c_1 = 0$ and $c_2 = 3/G$. Considering the action \eqref{bms-PGB} of our current work, we identify that the first part of the action coming from $p(\theta)$ is what we have obtained via dimensional reduction. This matching can be obtained by fixing the parameters $\tau_{\infty}$ and $\phi_r$ to appropriate values, which were arbitrary in the context of our earlier work. \\

Therefore, we have explicitly given an example of `celestial holography' for 3D gravity with zero cosmological constant. Part of this theory was obtained via holographic reduction as well. Our theory correctly reproduces the semiclassical entropy of 3D FSC, which further validates the claim. We would like to compute the logarithmic corrections to FSC entropy \cite{Bagchi:2013qva} from the 1D theory in a future work. \\

We are interested in computing the complete one-loop path integral of this 1D theory, which can potentially capture the 3D gravity path integral around an FSC saddle. The 1D theory is a Schwarzian coupled to a Lagrange multiplier field, thus it would be intriguing to analyze whether there is a connection between this 1D effective theory and BMS$_3$ invariant matrix models \cite{Bhattacharjee:2021zju} along the line of Schwarzian and hermitian matrix model duality \cite{Saad:2019lba}. It would be interesting to properly compare the 2D holographic notion and our 1D approach with other observables. In this context, the geometric action on BMS$_3$ coadjoint orbits and their connection to Liouville-like theories could be helpful \cite{Barnich:2017jgw, Merbis:2019wgk}.

\acknowledgments
We would like to thank Arjun Bagchi, Nabamita Banerjee, Geoffrey Comp\`{e}re, Suvankar Dutta, Marc Geiller, Dileep Jatkar, Jun Nian, and Max Riegler for useful discussions regarding various concepts related to this work. We are grateful to the people of India for their continuous support towards research in basic sciences.

\bibliography{CC.bib}
\end{document}